\RequirePackage{fix-cm}
\documentclass[smallcondensed]{svjour3} % onecolumn (ditto)
\smartqed % flush right qed marks, e.g. at end of proof
\usepackage{graphicx}
\begin{document}

\title{Dark Matter in Galaxies: evidences and challenges}

%\titlerunning{Short form of title} % if too long for running head

\author{ Paolo Salucci}


\institute{P Salucci \at
 SISSA \\
 Tel.: +39-40-3787520\\
 Fax: +39-40-3787528\\
 \email{salucci@sissa.it} 
}
 
\date{Received: date / Accepted: date}

\maketitle

\begin{abstract}
The evidence of the phenomenon for which, in galaxies, the gravitating mass is distributed differently than the luminous mass,
increases as new data become available. Furthermore, this discrepancy is well structured and it depends on the magnitude and
the compactness of the galaxy and on the radius, in units of its luminous size $R_{opt}$, where the measure is performed. 
For the disk systems with $-13\geq M_I\geq -24$ all this leads to an amazing scenario, revealed by the investigation
of individual and coadded rotation curves, according
to which, the circular velocity follows, from their centers out to their virial radii, an universal profile
$V_{URC} (r/R_{opt}, M_I)$ function only of the properties of the luminous mass component. Moreover, from the 
Universal Rotation Curve, so as from many individual high quality RCs, we
discover that, in the innermost regions of galaxies, the DM halo density profiles are very shallow. Finally, the disk mass,
the central halo
density and its core radius, come out all related to each other and
to two properties of the distribution of light in galaxies: the luminosity and the compactness. This phenomenology, being
absent in
the simplest $\Lambda CDM$ Cosmology scenario, poses serious challenges to the latter or, alternatively, it requires a
substantial and tuned involvement of baryons in the formation of the galactic halos. On the other side, the URC helps to
explain the two-accelerations relationship found by McGaugh et al 2016, in terms of only well known astrophysical
processes, acting in a standard DM halos + luminous disks scenario.

\end{abstract}

\section{Introduction}

 The presence of huge content of invisible matter in and around spiral galaxies, distributed differently from stars and gas,
is well determined from optical and $21$ cm rotation curves (RCs) \cite{rubin80,Bosma81}. The extra mass component
becomes progressively more abundant {\it i)} at outer radii and {\it ii)} at a same radius, in the less luminous galaxies (
\cite{ps88}). 
The total gravitational potential of spirals $\phi_{tot}$ includes different components: $\phi_{tot}=\phi_b+\phi_d+\phi_{HI}+
\phi_{DM}$ namely, the bulge component, the stellar disk component, the HI disk component and finally the Dark Matter one.
$\Phi_{tot}$ is related to the galaxy's circular velocity by:
\begin{equation}
V^2(r)=r\frac{d}{dr}\phi_{tot}=V^2_b +V^2_d+V^2_{HI} + V^2_{DM}
\end{equation}
with all the R.H.S. terms function of radius $r$. The Poisson equation relates the surface/spatial densities to
the corresponding gravitational potentials. 
Then, the velocity fields $V_i$ are the solutions of the four separated Equations:
\begin{equation}
 \nabla^2 \Phi_i= 4 \pi G \rho_i
\end{equation}

\begin{figure}
\begin{centering}
 \includegraphics[width=0.91\textwidth]{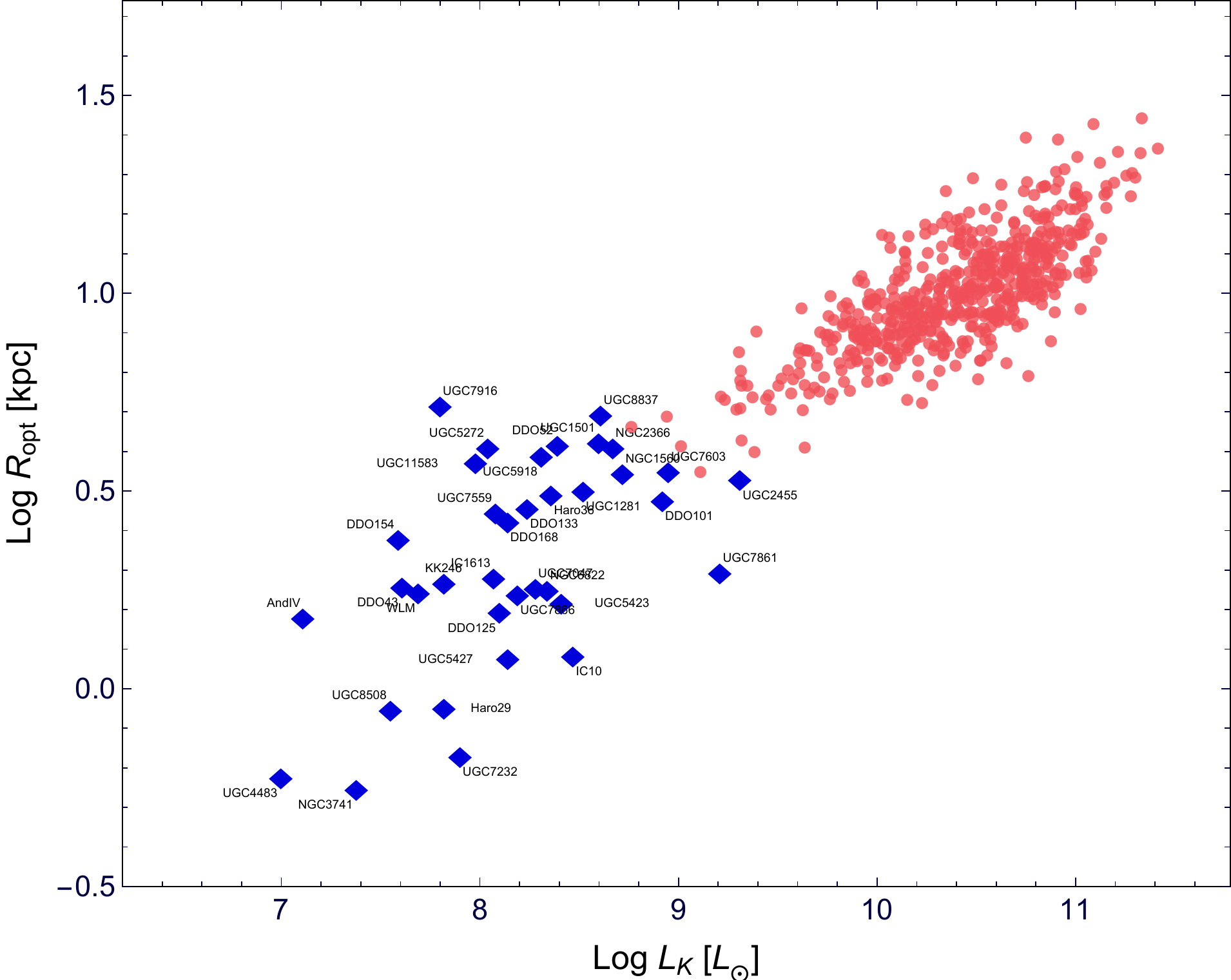}
\caption{The optical radius $R_{opt}$ vs luminosity for spirals (red circles) and dwarf disks (blue diamonds)}
\end{centering}
\end{figure}

where the index $_i$ defines the 4 components of the total density: $ \rho_{b}(r)$, $\mu_{d}(r) \delta (z)$, $\mu_{HI}(r)
\delta (z)$, $\rho_{DM}(r)$, with $\delta(z)$ the Kronecker function and $z$ the cylindrical coordinate.We can generally
assume that the stellar surface density $\Sigma_d(r)$
is proportional to $\mu_d(r)$, well measured by CCD infrared photometry, leading to the well-known Freeman
exponential thin disk profile \cite{freeman} 
\begin{equation}
\Sigma_{d}(r)=\frac{M_{D}}{2 \pi R_{D}^{2}}\: e^{-r/R_{D}},
\end{equation}
where $M_D$ is the disk mass and $R_D$ the scale length. $R_{opt}\equiv 3.2 R_D$, the radius that encloses $ 83\%$ of the
total 
galaxy light, is usually adopted as the optical size of the galaxies. In Spirals, the two above quantities
are well correlated (see Fig. (1) and \cite{Tonini}):
\begin{equation}
\log\left(\frac{R_D}{\mbox{kpc}}\right) = 0.633+0.379\log\left( \frac{M_D}{10^{11}M_{\odot}}\right) 
+ 0.069\left(\log \frac{M_D}{10^{11}M_{\odot}} \right)^2,
\end{equation}

From Eqs (2-3) and with $y \equiv r/R_D$ we have: 

\begin{equation}
V_{d}^{2}(y)=\frac{G M_{D}}{2R_{D}} y^{2}B\left(\frac{y}{2}\right)
\end{equation}

where $y\equiv r/R_{D}$, $G$ is the gravitational constant $B=I_{0}K_{0}-I_{1}K_{1}$ a combination of Bessel functions
evaluated a $1/2 ~y$, \cite{freeman}.Let us stress that to adopt directly in Eq(2) the measured surface brightness, rather
than
its fitting function in Eq(3), changes no result of this work 

\begin{figure} 
\begin{centering}
\includegraphics[width=1.05\textwidth]{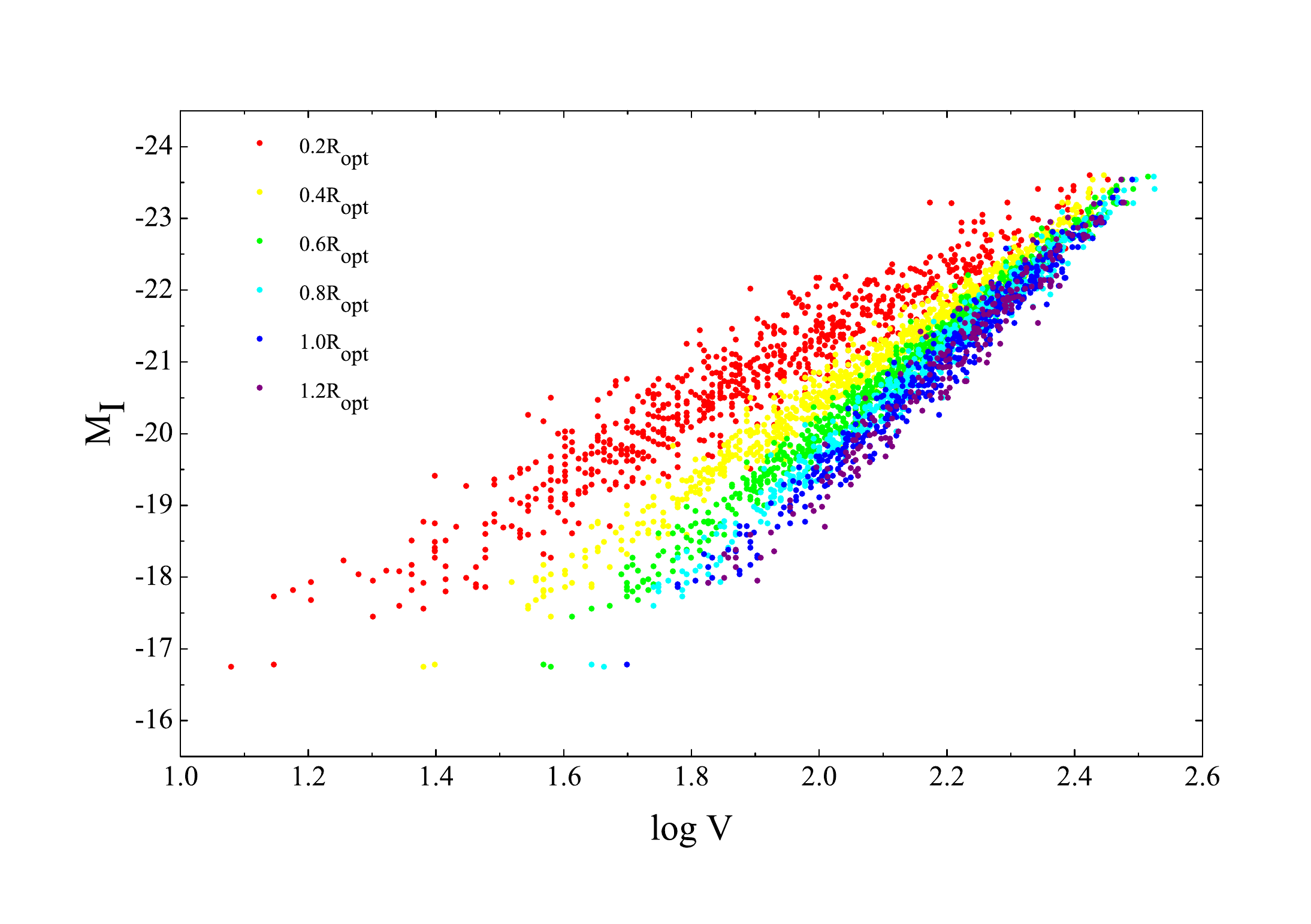}
\caption{The Radial Tully-Fisher: the ensemble of the relations at different radii measured in units of $R_{opt}$.} 
\end{centering}
\end{figure} 

The surface density of the HI disk $\Sigma_{HI}(r)$ is directly derived by 21 cm HI flux measurements and it can be
approximately represented by 
$1/9 ~\Sigma_{d}(r/(3~R_D)) (M_{HI}/M_D)$
\cite{Tonini}, then: 

\begin{equation}
V_{HI}^2(y)= \frac { M_{HI}} {9 ~M_D} ~ V_d^2 (y/3) 
\end{equation}

Stellar Bulges are important contributors of the total galaxy mass only for early Hubble Types objects that are non-considered
in this work. 

Let us to introduce here, for any galaxy, the virial radius $R_{vir}$ and the virial halo
mass $M_{vir}$ related by: $ M_{vir} \simeq 100 \rho_c R_{vir}^3$ and $\rho_c$ is the mean density of the Universe:
$\rho_c=1 \times 10^{-29}\ g/cm^3$

The rotation curves of spirals show properties and an high degree of universality that cannot be explained by their baryonic
matter content:

$\bullet$ {\bf Amplitudes}. At any radius $R_n$, measured in units of $R_{opt}$ such that: $R_n \equiv
(n/5)R_{opt}$, ($ n=1, 7 $), there is a tight relationship between the local rotation velocity $V_n \equiv V(R_n)$ and the
total
galaxy magnitude $M_I$ \cite{ys}: 

\begin{equation}
M_I = a_n \log V_n + b_n 
\end{equation} 

The ensemble of relationships is shown in Fig (2); their r.m.s. scatter is always very small ($<
0.3$ magnitudes) and for $n=3$ reaches a
minimum of $0.12 $ magnitudes (\cite{ys}). This baffling result indicates that, in average, the I-magnitude is able to
predict, in any galaxy and at any radius, the value of the circular velocity within a $5\%$ uncertainty. Moreover, the
evident increase of $a_n$ with $n$, see Fig. (2) provides us with precious information on the mass distribution in Spirals
\cite{ys}. 

\begin{figure}
\begin{centering}
\includegraphics[width=0.8\textwidth]{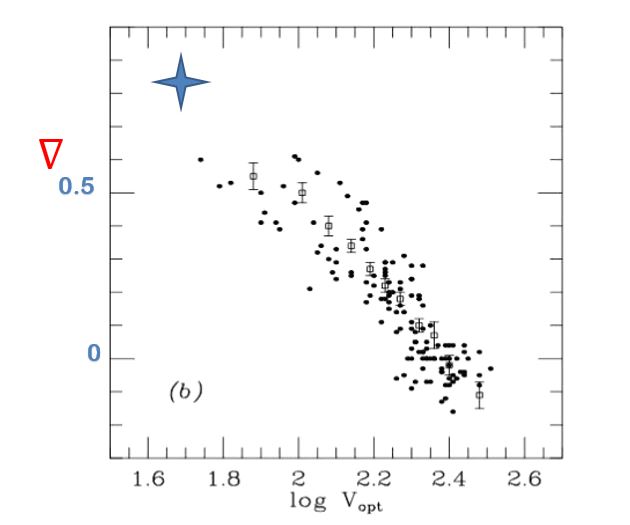}
\caption{The rotation curve slope $\nabla \equiv dlog \ V/dlog \ r$ at $R_{opt}$ as a function of $log \ V_{opt}$. The cross
indicates the region of {\bf dd} galaxies. Also shown the data from the 11 coadded {\it(open circles with errorbars)} and
131 individual {\it (filled circles)} RCs of \cite{pss}. }
\end{centering}
\end{figure}

 $\bullet$ {\bf Slopes} $\nabla$, the logarithmic slope of the circular velocity at $R_{opt} $ emerges as a tight 
function of
$V_{opt}$ and of galaxy magnitude (see Fig(4) and also \cite{pss} ). One finds: $ -0.3 \leq \nabla \leq 1$ (see Fig. (3)).
Let
us also stress, that the quantity $\nabla$ takes, in disk systems, all the values allowed in Newtonian Gravity, from -0.5
(Keplerian regime)
to 1 (solid body regime), falsifying so, the paradigm of ``flat rotation curves'' according to which, in great prevalence,
one should find: $\nabla= 0$.

\section {The Universal Rotation Curve of Spirals} 

 We can represent the rotation curves of late types Spirals by means of the Universal Rotation Curve (URC) pioneered in
\cite{rubin2,p91} and set by \cite{pss} and \cite{s07}. The first step of the investigation of the spiral kinematics is the
acquisition of 11 coadded rotation curves
$V_{coadd}(r/R_{opt}, M_I)$ that are obtained, by binning and averaging in {\it a)} magnitude and {\it b)} normalized
radius
$x\equiv R/R_{opt}$,
$967$ extended and high quality rotation curves of late type spirals (published in \cite{ps95}).
 \begin{figure}
\begin{centering}
\includegraphics[width=1.0 \textwidth ]{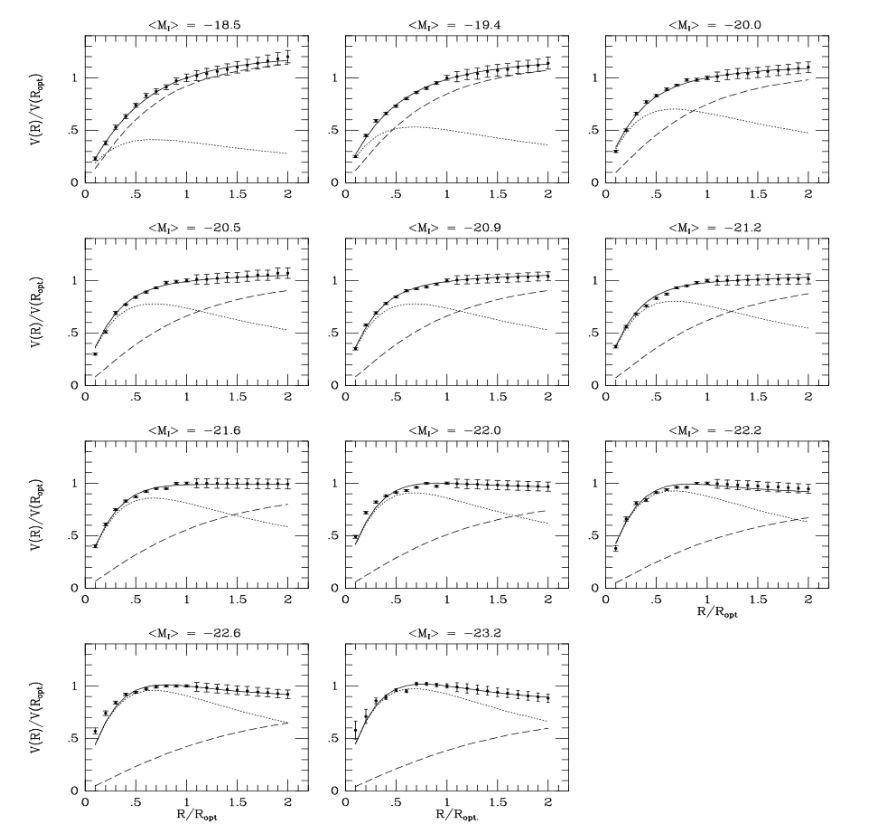}
\caption{The coadded RCs ({\it points}) and the Universal Rotation Curve of Spirals out to $6.4\ R_D$ ({\it lines}). Also
shown
the dark ({\it dashed}) and the luminous ({\it pointed}) velocity components.} 
\end{centering}
 \end{figure}

 These 11 coadded RCs ({\it points} with errorbars in Fig(4)) extend out to $\simeq 2 \ R_{opt}$ and represent the full
kinematics
of spirals, whose I-magnitude range is $-16.3 < {\rm M}_I< -23.4$. They
lead to the Universal Rotation Curve (URC): i.e. a velocity model $V_{URC} (r/R_{opt}, M_I)$ function of radius and of 
luminosity, that well fits the $V_{coadd}(r/R_{opt}, M_I)$ data (see Fig(5) and \cite{pss,s07}). 

The URC is, therefore, a specific proper function of
normalized radius, which, tuned by few parameters, namely the galaxy luminosity, well fits the coadded and individual rotation
curves representing the RCs of more than 100k local spirals of different luminosity and Hubble type. 

 In detail, in the simplest version, the URC has two velocity components, one from the stellar disk and the other from the
dark halo: 
\begin{equation}
V^2_{URC}(x,M_I) = V^2_{URCd}(x,M_I) + V^2_{URCh}(x,M_I) 
\end{equation}
The first component is the standard Freeman disk of Eq. (4),
\begin{equation}
V_{URCd}^{2}(y)=\frac{G M_{D}}{2R_{D}} y^{2}B\left(\frac{y}{2}\right)
\end{equation}

 The
second is the Burkert halo velocity profile, proposed by \cite{SB} to represent the DM density in halos around galaxies of
any magnitude or Hubble Type:

\begin{figure}
\begin{centering}
\includegraphics[width=0.87\textwidth ]{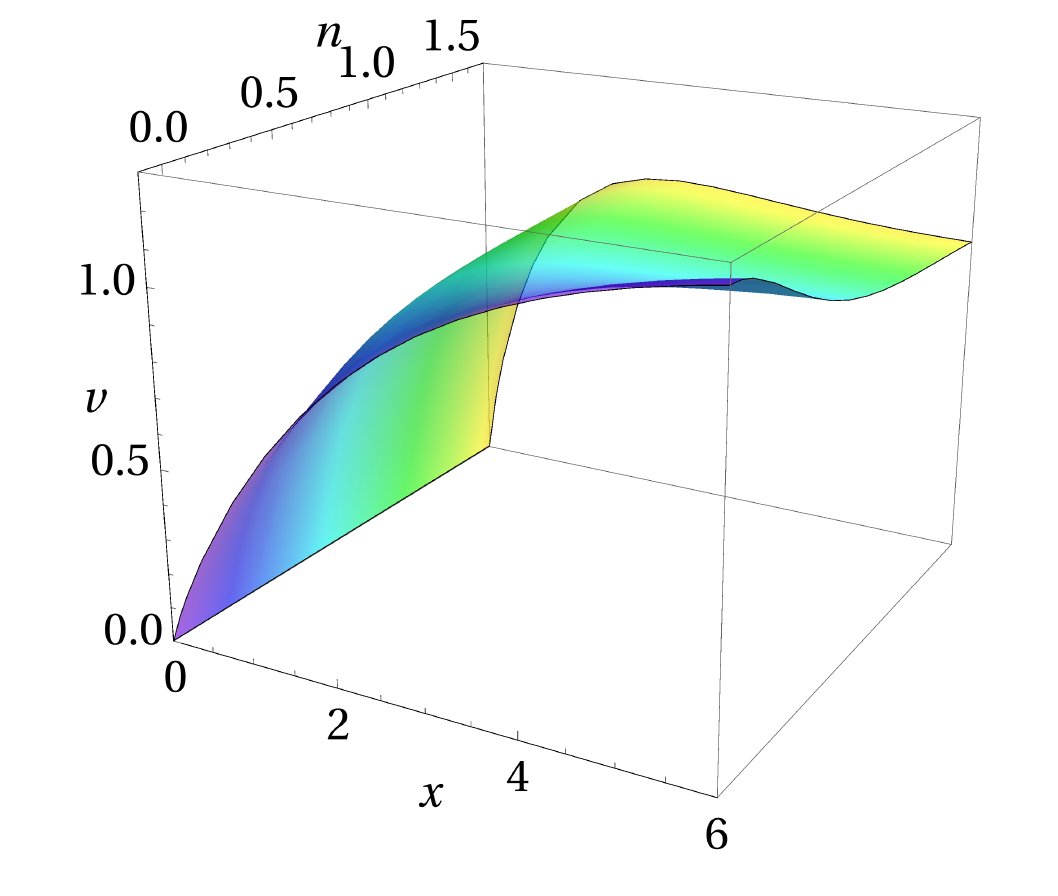}
\caption{The URC out to 2 $R_{opt}$. We have: $ n=log \ M_{vir} -11$, $v\equiv V(r/R_{opt})/V(R_{opt})$} 
\end{centering}
\end{figure}
 
\begin{equation}
\rho (r)={\rho_0\, r_0^3 \over (r+r_0)\,(r^2+r_0^2)}~,\label{BH}
\end{equation}
\begin{equation}
V^2_{URCh}(r) =6.4\frac{\rho_0r_0^3}{r}\big( \ln (1+\frac{r}{r_0})-\arctan \frac{r}{r_0}) +\frac{1}{2}\ln ( 1+\frac{r^2}
{r_0^2})\big). 
\end{equation}

where $\rho_0$ and $r_0$ are, respectively, the DM central density and its core radius. The URC velocity model has then three 
parameters: $M_D, \rho_0, r_0$ that are obtained by best-fitting the 11 coadded rotation
curves (that represent the kinematics of the whole family of normal Spirals). The fit is excellent (see Fig.(4) and
\cite{pss}). The reduced $\chi^2 <
1 $ and the uncertainties on the parameters are about $15\%$. Each parameter is related to all the others see Fig.(6) and they
all are dependent on luminosity \cite{s07}. As far as the halo virial mass $M_{vir}$ we have: ($M_s=3
\times 10^{11} M_\odot$):
$
M_D = 2.3\ 10^{10} (M_{vir}/M_s)^{3.1}/(1 + (M_{vir}/M_s)^{2.2})
$

The mass distribution in Spirals as resulting from the URC (see \cite{s07} for
details) has some specific charactheritics. At any normalized radius $x$, objects with lower
luminosity have a larger dark-to-stellar mass ratio. Moreover, spirals have a radius, whose size increases from $0.5 R_D$ to
$3 R_D $ with galaxy
luminosity, inside which 
 the baryonic matter fully accounts for the rotation curve and outside which, a dark component is needed 
to justify the RC profile (see Fig (4)) in (\cite{pss}) Let us notice that the latter is the correct enunciation of the
wellknown ``maximum disk hypothesis.

\begin{figure}
\begin{centering}
\includegraphics[width=0.82 \textwidth ]{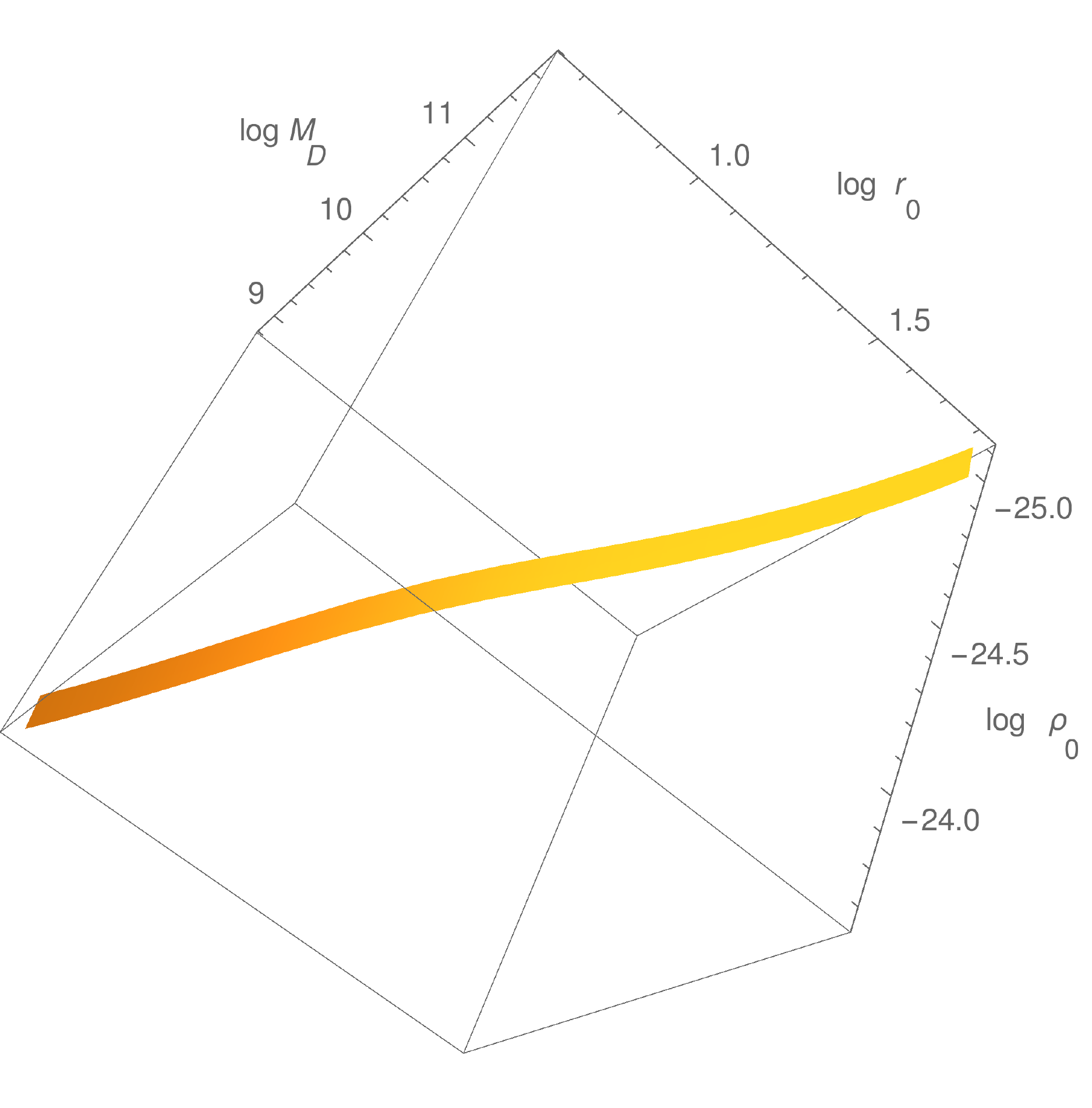}
\caption{The relationship among the URC parameters. Units: $M_D$ in $M_\odot$, $\rho_0$ in $g/cm^3$ and $r_0$ in kpc. } 
\end{centering}
\label{fig:new}
\end{figure}

 Of particular importance is the quantity $\mu_{0D}\equiv \rho_0 r_0$,
proportional to the halo central surface density, that results constant in objects of any magnitudes and Hubble Type, as
pioneered by
\cite{Kormendy,donato04}:
\begin{equation}
 log \ \frac{\mu_{0D}}{\rm M_{\odot} pc^{-2}} = 2.2 \pm 0.25
 \end{equation}
This relationship is supported by independent work \cite{spano08} and it can be considered as a portal leading
to the nature itself of the dark matter \cite{donato09} (see also Section 4). 
 
By means of a number of very extended RCs 
and of virial velocities $V_{vir}\equiv (G M_{vir}/R_{vir})^{1/2}$ obtained by means of the abundance matching method,
\cite{shankar06}, it is
possible to determine with accuracy the halo mass \cite{s07} around a galaxy of magnitude $M_I$ and therefore to extend the
URC
out to galaxie's
virial radii (see Fig(7))
\begin{figure}
\centering
\includegraphics[width=0.9\textwidth ]{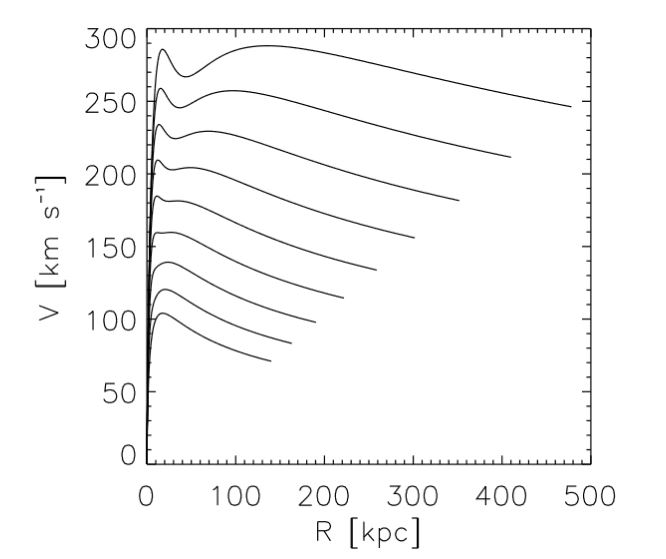}
\caption{ The Universal Rotation Curve out to $R_{vir}$ in physical units, see \cite{s07} for details} 
\label{fig:new}
\end{figure}

As result, in Spirals, the halo mass range is: $3 \times 10^{10} M_{\odot} \leq M_{vir}\leq 3\times10^{13} M_{\odot} $. The
stellar to halo fraction $M_D/M_{vir}$ ranges between $7\times 10^{-3}$ to $5\times 10^{-2}$, \cite{evoli}, values much
smaller than the cosmological one of $\Omega_b/\Omega_{matter} \simeq
1/6 $,

\section{ Cuspy or Cored Dark Matter Halos in disk systems} 

The lack, in the DM halo density of Spirals, of the inner cuspiness predicted by N-Body simulations in (the simplest version
of) the $\Lambda$ Cold Dark Matter scenario\cite{nfw96} 
is a crucial evidence for the fields of Cosmology and in Astroparticle.
In fact, in such scenario, the DM halo spatial density is universal and it is well reproduced by one-parameter radial
profile \cite{nfw96}:

\begin{equation}
\rho_{NFW}(r) = \frac{\rho_s}{(r/r_s)\left(1+r/r_s\right)^2},
\label{eq:nfw}
\end{equation}

where $r_s$ is a characteristic inner radius, and $\rho_s$ the corresponding density. It is clear that the NFW halo density
diverges at the origin as $r^{-1}$. From Eq. (13) we get:

 \begin{equation}
V^2_{NFW}(r) = M_{vir} \frac {( ln \ (1+r/r_s)-(r/r_s)/(1+r/r_s)} {ln (1+c)-c/(1+c)}/r
\end{equation}
with: $
R_{vir} / r_s \simeq 9.7 \left( \frac{M_{vir}}{10^{12}M_{\odot}} \right)^{-0.13}$.

Since their emergence in numerical
simulations, cuspy density profiles were claimed to disagree with the DM density profiles
detected around dwarf spirals (e.g. \cite{moore94}). However, strong concerns were raised on whether the evidence provided 
was biased by observational systematics. 

\begin{figure}
\begin{centering}
\includegraphics[width=0.75\textwidth]{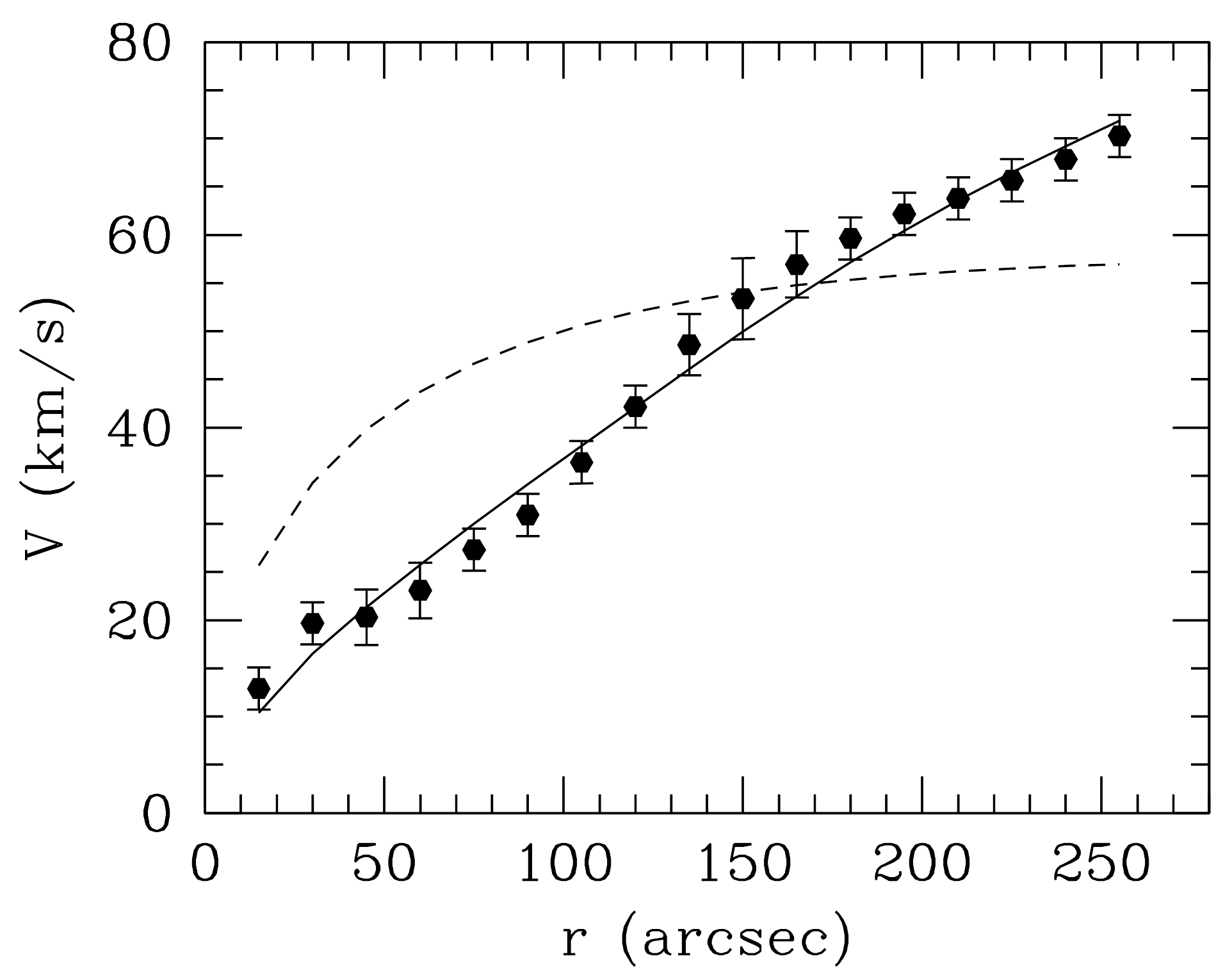}
\caption{Rotation curve of DDO 47 ({\it points}) vs models. Burkert halo + stellar/HI disks ({\it solid line}). NFW halo +
stellar/HI disks ({\it dashed line}) (see \cite{gentile05})}.
\end{centering}
\end{figure}
A solution of the cusp-core controversy came from a careful modelling of 2D, high quality, extended rotation curves
\cite{gentile04}. As result of this strategy, no cuspy behavior in the DM density has been found (e.g. \cite{deblok2,SD}).
Presently, none of the ~100 most suitable and high quality 
RC can be satisfactory reproduced by a NFW halo + stellar HI/disks velocity model. In virtually all the cases, instead, the
cored
model fits well the RC with reasonable values for the free parameters. As a test case, we consider the
nearby dwarf spiral galaxy DDO 47 (see Fig.(8)). Its RC modelling finds that in this galaxy 
the dark halo density must have a core of $ \sim 7$ kpc and a central density of $\rho_0 = 1.4 \times 10^{-24}$ g cm$^{-3}$
 The NFW halo profile, instead, is totally unable to fit the RC see Fig (8).

It is important to stress that the cusped
halo distributions do fail, not (only) because they fit poorly the RCs, but also because (see e.g. \cite{gentile04}):

 $\bullet$ they often imply implausibly values for the stellar mass-to-light ratio and/or the halo mass

 $\bullet$ they often do not follow the $\Lambda CDM$ concentration vs halo mass relationship
 
Moreover, direct investigations of RCs have also ruled out the possibility that the detection of a cored distribution could
be a
mirage arisen by not-axysimmetric motions in the galaxy affecting the rotation curve (e.g. \cite{gentile05}).

Finally, we notice that there are {\it model independent} evidences for cored DM halo density distributions.
Salucci, (2001) \cite{Salucci01} derived, for 140 spirals of different luminosity, $\nabla_h$ the logarithmic
gradient of the halo contribution to the circular velocity ad the edge of the stellar disk: $\nabla_h\equiv \frac{d \log
V_h(r)}{d \log r}$ evaluated at $R_{opt}$ (see Fig. (9)). He found: $ \nabla_h\simeq 0.9 $ in all galaxies, i.e. a value
inconsistent with the predictions by the NFW density profile. For a large sample of
Low Surface Brightness galaxies, a very similar result was obtained by \cite{deblok1}.
 \begin{figure}
\begin{centering}
\includegraphics[width=.93\textwidth]{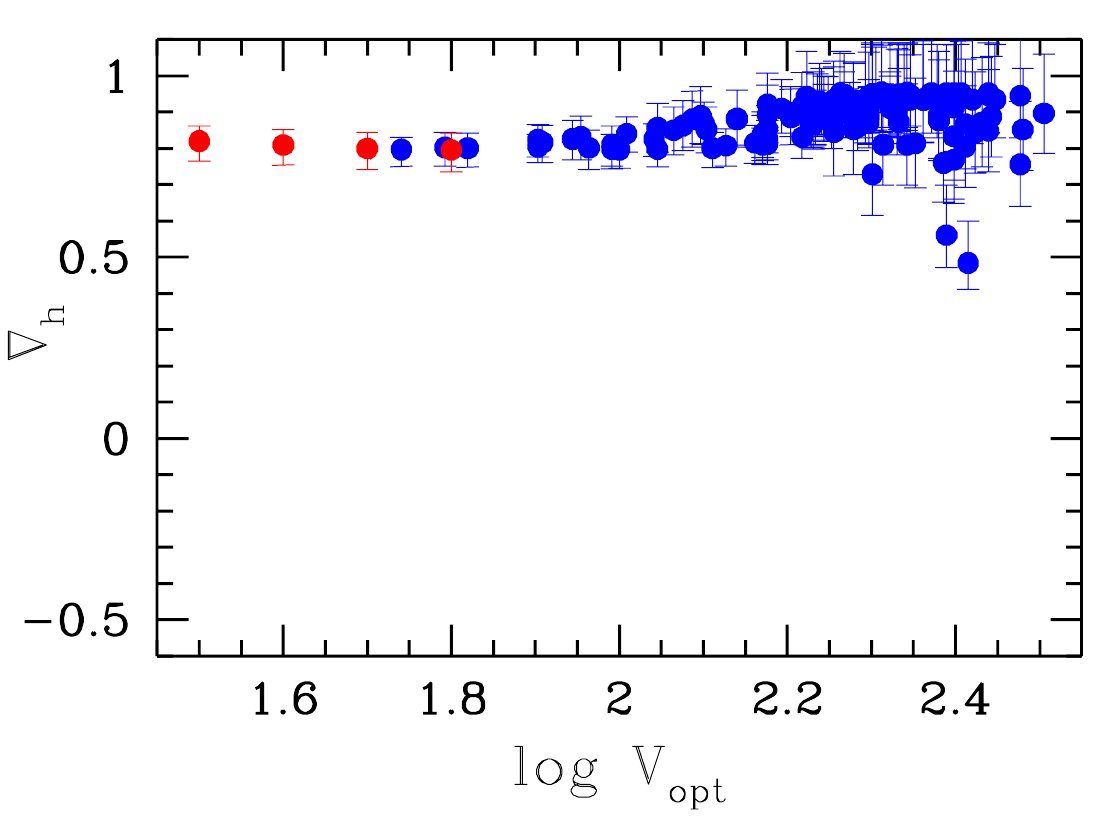}
\caption{ $\nabla_h$, the DM halo velocity slope at $R_{opt}$ in Spirals ({\it blue}) and in {\bf dd} ({\it red}).  The NFW
halo profiles have: $\nabla_{h,NFW} \simeq 0.2 $}
\end{centering}
\end{figure} 

 \begin{figure}
\begin{centering}

\includegraphics[width=0.847\textwidth]{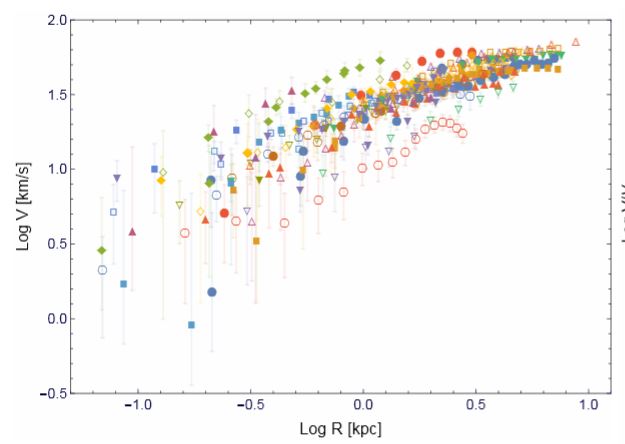}
\vskip 0.1cm
\advance\rightskip 0.1truecm
\includegraphics[width=0.816\textwidth]{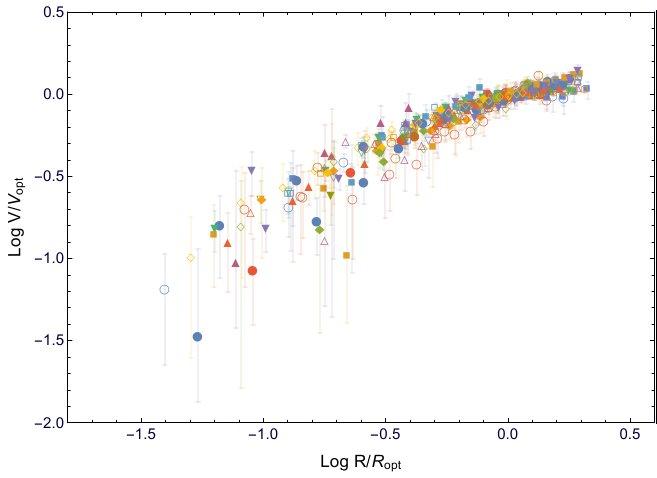}

 \end{centering}
\caption{Individual RCs of {\bf dd}. In physical units ({\it top panel}). After the $R_{opt}$ and
$V_{opt} $ double normalization ({\it bottom panel}). Each galaxy has its own color-shape code, see \cite{ks}.}
\end{figure}

\section{The URC of dd galaxies}

Karukes and Salucci, 2016 (\cite {ks}) selected a sample of 36 dwarf discs from the Local Volume Catalog, 
which is $\sim 70$ per cent complete down to $M_{B}\approx-14$ and out to 11 Mpc. The objects are bulgeless systems in which
rotation, corrected for
the pressure support, balances the gravitational force. Morphologically, they include gas-rich dwarfs
star-forming at a relatively-low rate, and starbursting blue compact dwarfs (BCD). Hereafter, for simplicity, we call them 
dwarf disks ({\bf dd}) that, for the DM investigation, is their principal characteristic.
They have a Freeman surface luminosity profile but, differently from spirals, their $L_K$ vs $R_{opt}$
relationship has a very large scatter, see Fig.(1).
The disc length scales $R_D$ of the galaxies in the sample are known within 15 per
cent uncertainty; their RCs are symmetric, smooth with small r.m.s. and extend out to $\sim$ 3 $R_D$ 

In this sample that reaches 5 magnitudes down with respect to the least luminous spirals (\cite{pss}), 
the magnitudes, disc lenght scales and optical velocities intervals are:
$$
 -19 <M_{I} < -13, \ \ 0.18 \ kpc <R_{D} < 1.63 \ kpc, \ \ 17 \ km/s\ < V_{opt} < 61 \ km/s
$$

 The average optical radius and optical velocity of the sample are:$ \langle R_{opt} \rangle$ and $ \langle V_{opt} \rangle$
are:
$2.5
\ kpc$, $40.0 \ km/s$, respectively. 

\begin{figure}
 \begin{centering}
\includegraphics[width=1 \textwidth]{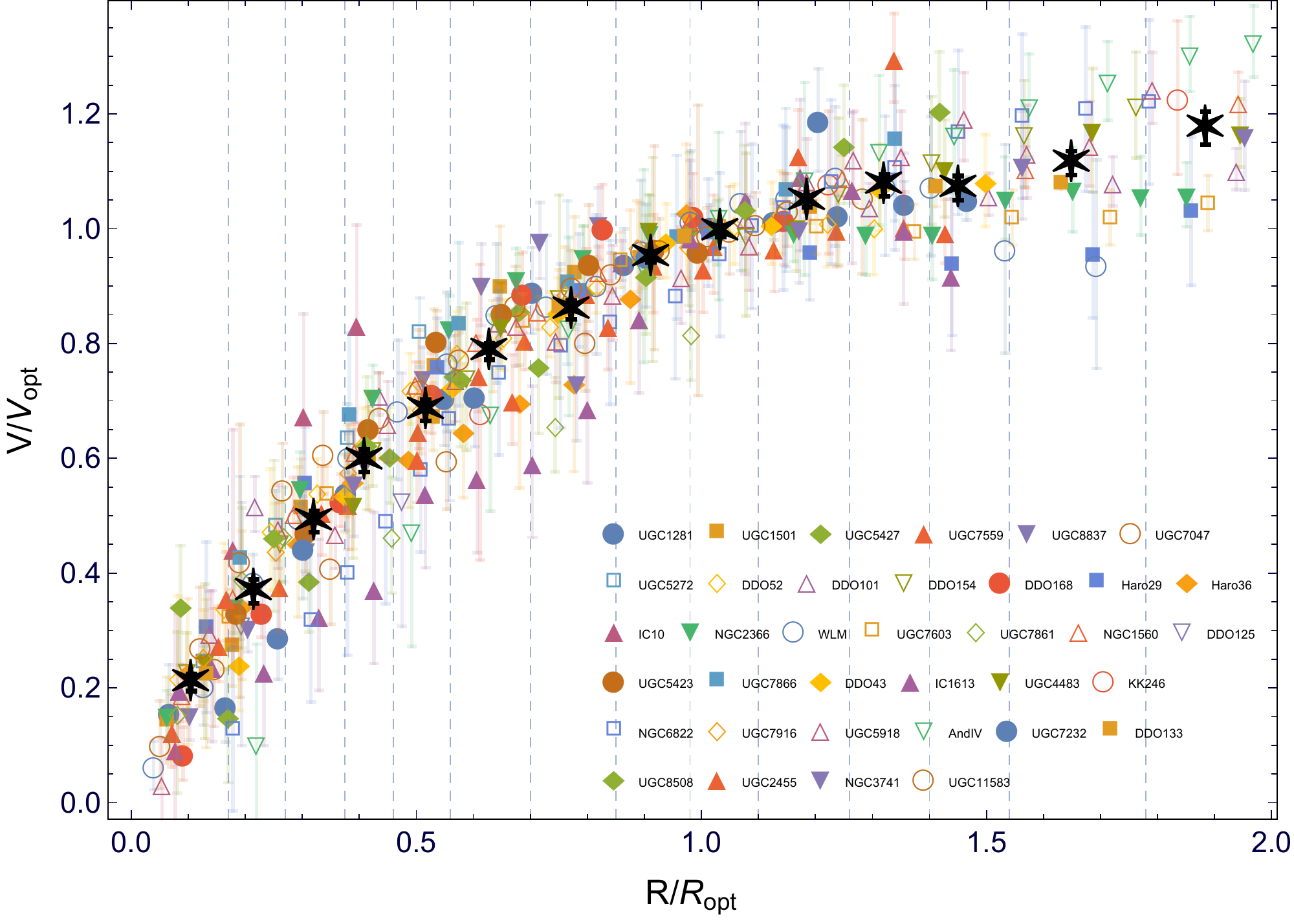}
\caption{Individual RCs normalized to $R_{opt}and V_{opt}$. Also shown the coadded curve ({\it black stars}). The bins are
indicated by vertical dashed grey lines.}
\label{fig3}	
\end{centering}
\end{figure}

We plot, in the log-log scales, the 36 RCs of the {\bf dd} sample expressed in physical units (Fig. 10 {\it top}). Contrary to
the RCs
of normal spirals \cite {pss}, each {\bf dd} rotation curve has a quite different
shape, see also \cite{oman15}. Although all curves increase with radius, this occurs, for each galaxy, at a
very different pace. This behavior is related to the very large scatter of the $R_{opt}$ vs $L_K$ relationship, evident 
in Fig. (1) and absent in normal spirals. In fact, let us double normalize the 36
$V(r)$ to their $R_{opt}$
and $V_{opt}$ values and then derive the quantity $v(x)\equiv V(r/R_{opt})/V(R_{opt})$. This quantity has an unique profile
for all objects 
(see Fig 10, {\it bottom}): the double normalization of the RC has eliminated most of their original diversity.
All the RCs of the Sample are then placed in a same luminosity bin (see Fig (11)).

We coadd the double normalized 350 velocity data by setting 14 radial bins centred at $r_i$ ($i=1,14)$ 
 Every bin has a number of data from a maximum of 68 to a minimum of 14. Then, by averaging the data in each radial
bin $i$ we derive $v_i $, the coadded (double normalized) rotation velocity at $r_i$: $v_i = 
V(r_i/R_{opt})/V(R_{opt})$, their r.m.s. $\sigma_i$ and residuals $dv_{ij} = v_{ij}-v_i$. The two latter quantities result
always very small
( \cite{ks}) as required by the URC paradigm. 
The individual RC's ({\it different colors}) and the coadded one ({\it big stars}) are shown in Fig(11).

 The {\bf dd} luminosity range is as large as that of spirals (PSS).
 However, differently from them, their rotation curves show all the same (two-normalized) profile and precisely that of the
coadded
rotation curve of the least luminous normal spirals, with $ M_{I} \simeq -18.5$ (see \cite{pss,ks}). 
From this magnitude down, in all disk systems, the RC profile becomes a solid-body like: $V(r)\propto r$ and the stellar disc
contribution disappears from the kinematics.

We build the coadded {\it fiducial} rotation curve: first, for simplicity,
we rescale the double normalized velocities $v_i$ to the average values of the sample: $ \langle
V_{opt}\rangle$ and $\langle R_{opt}\rangle$, 40.0 km/s and 2.5 kpc. So, $ \langle
V_{i}\rangle=v_{i} \langle V_{opt}\rangle$
and $\langle R_{i}\rangle=r_{i} \langle R_{opt}\rangle$. The coadded {\it fiducial} {\bf dd} RC extends out to 1.9 $\langle
R_{opt}\rangle$ and it has uncertainties of $\sim 5 \%$ (see Fig(11)).
 
The URC, in the present case: $V_{URC}(x, \langle V_{opt}\rangle)$, is the {\it halo + disks velocity model} that fits the
fiducial {\bf dd} coadded RC, shown in Fig. (11) as big stars and in Fig (12) as filled circles with errorbars. It consists
into the sum,
in quadrature, of three terms: $V_{URCd}$, $V_{URCHI}$, $V_{URCh}$ that describe the stellar disc, the HI disc and the dark
halo contributions. 
The {\bf dd} galaxies in the sample have all Freeman surface density profile 
\cite{freeman}:
that leads to the velocity term given in Eq.(5). For the HI component we have: 
 $V_{URCHI}^2 (r)= 1/9 \ V_d^2(x/3)~ M_{HI}/M_D$ \cite{Tonini}. Moreover: $\langle M_{HI} \rangle = 1.7 \times 10^8
M_{\odot}$ (see \cite{ys}). 

For the DM halo we use alternatively the Burkert and the NFW profiles introduced in the previous sections.
The URC model fits very successfully the fiducial RC (see Fig.(12)) $\chi^2_{reduced} <1$ and the best fit values
of the parameters are:

 \begin{figure}
 \begin{centering}
\includegraphics[height=9.7cm,width=11.9cm]{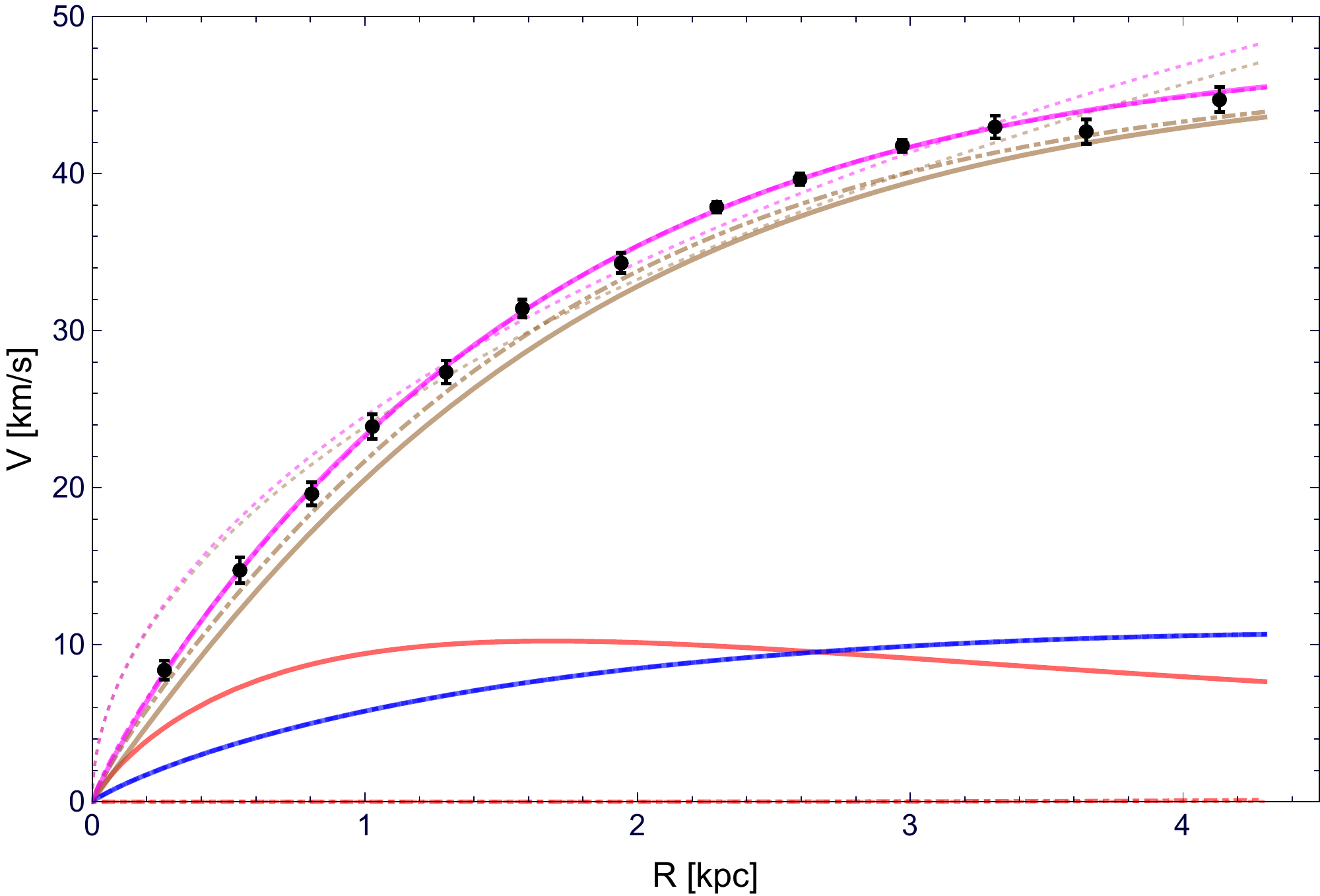}
\caption{The coadded {\bf dd} RC ({\it filled circles}) fitted by the URC ({\it pink solid line}) and by NFW ({\it pink
dashed line}) profiles. Also shown the relative contributions by stellar disc ({\it red line}), HI disk
({\it blue line}) and dark halo ({\it brown line})}
\end{centering}
\end{figure}

 $$
 log\langle \rho_0\rangle = 7.55 \pm 0.04 \ \
\langle r_0 \rangle = 2.3 \pm 0.13 \ \ log\langle M_{D}\rangle = 7.7 \pm 0.15 
$$ where $(\rho_0,r_0, M_D)$ are in units of $(M_\odot/kpc^3, kpc, M_\odot)$. The resulting virial mass is $ \langle M_{vir}
\rangle=(1.38 \pm 0.05) \times 10^{10} M_{\odot}$.

Instead, the NFW profile fails to reproduce the coadded RC (see dashed lines in Fig (12)), $\chi^2_{reduced} \approx
12$ and the values of the best-fit parameters are
$$ 
log\langle M_{vir}\rangle = 11.7 \pm 0.9 \ \ \langle c\rangle = 4.7 \pm 3.2; \ \ log\langle
M_{D}\rangle = 2.5_{-2.5}^{+?} \ \
$$
where the two masses, in units of $log ~M_\odot$, are totally unrealistic. 

By connecting the best fitting values of the core radii with the corresponding stellar disk lenght-scales we have that also
{\bf dd} stay on the line:
\begin{equation}
 log \ r_0=0.47+1.38 \ log \ R_D 
\end{equation} 
found in Spirals (\cite{s07}).

Since: {\it a)} the double normalized coadded RC is very good fitted by the {\bf dd} URC and {\it b)} given the
strong
correlation of Eq.(14), it is possible to obtain from the double normalized URC $V(r/\langle R_{opt}\rangle)/\langle
V_{opt}\rangle$ the structural parameters of each galaxy of the sample.. The procedure is the following 
 (see also \cite{ks}): in each galaxy, we have, for both disk components: 
$$ 
\frac{M_{D,HI}}{V^2_{opt} R_{opt}} =\frac{\langle M_{D,HI}\rangle}{\langle
V^2_{opt}\rangle \langle R_{opt}\rangle}. 
$$

 We also assume that average value $\frac{\langle {V^2_D(R_{opt})}\rangle}{\langle V^2_{HI}(R_{opt})\rangle}\simeq1.1$ holds
in all the objects. 
Therefore, for each galaxy of the sample we have that:  e.g. the DM mass inside $R_{opt}$ takes the form:
$$
M_{DM}(R_{opt}) =(1-\alpha) V_{opt}^2 R_{opt} G^{-1}
$$,
where $M_{DM}$ is the DM mass inside the optical radius $R_{opt}$ and $\alpha$ is the fraction with which the baryonic matter
contributes to the total circular velocity. Then: 
\begin{equation}	
\alpha =\frac{\langle V_{HI}^2(R_{opt})\rangle+\langle V_{D}^2(R_{opt})\rangle}{\langle V_{tot}^2(R_{opt})\rangle}
=0.12.
\end {equation}

i.e. $\alpha$ is constant over the objects of the sample. Then, for each galaxy, by inserting in the above equations its
values for $R_{opt}, V_{opt}$, we obtain its dark and the luminous structural parameters. Thus, we realize that {\bf dd}
live in haloes with masses below
 $5 \times 10^{10} M_{\odot}$ and above $4 \times 10^{8} M_{\odot}$.

 \begin{figure}
 \begin{centering}
\includegraphics[width=1.\textwidth]{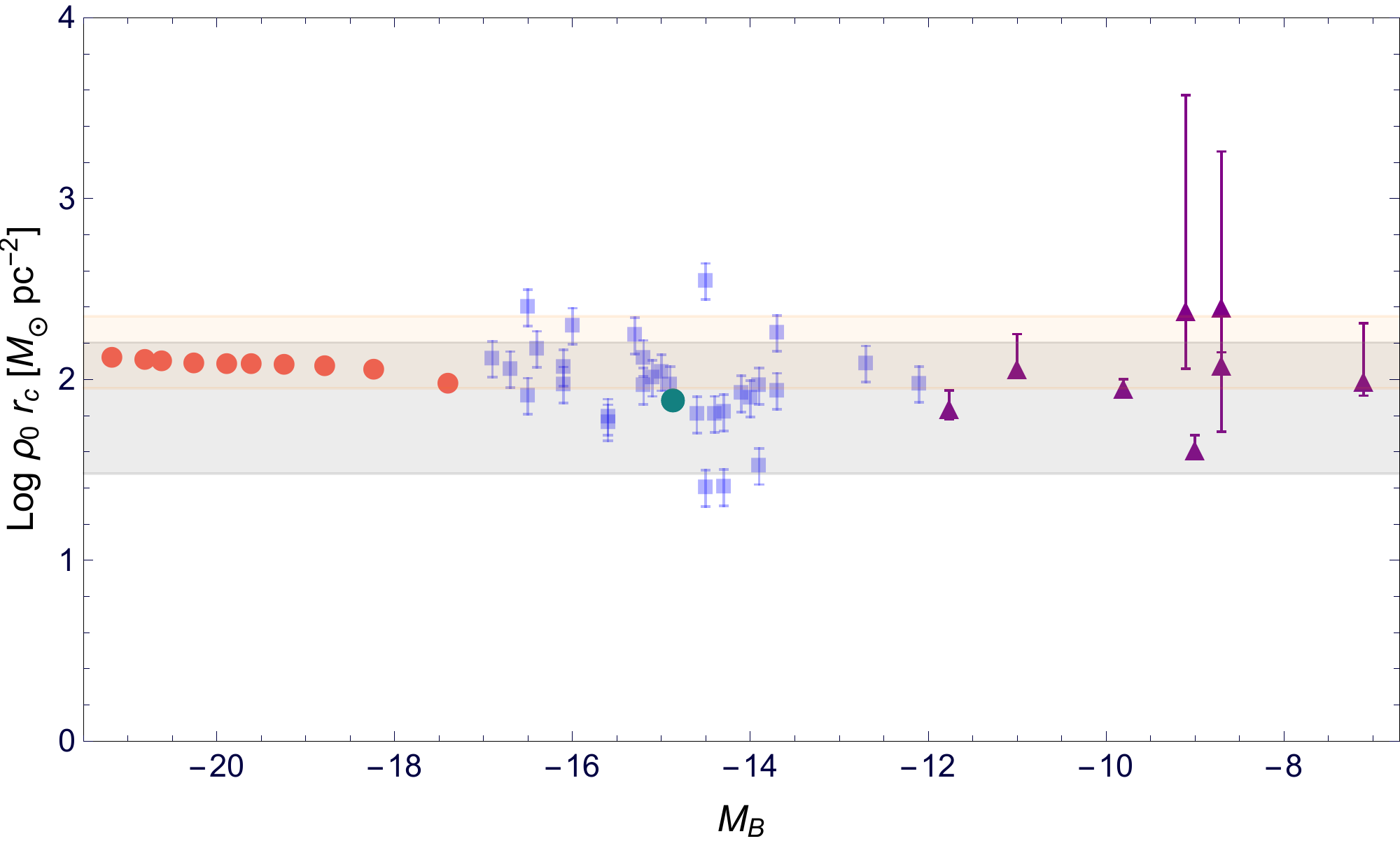}
\caption{$\rho_0r_c$ in units of $M_{\odot} pc^{-2}$ as a function of magnitude for galaxies of different Hubble
Types.
 Data come from: \cite{s07} URC of spirals ({\it red circles}); the scaling relation from \cite{donato09}
({\it orange shadowed area}); the Milky Way dSphs (purple triangles) \cite{salucci12}; {\bf dd} (blue squares-this work), 
the empirical relation: $\rho_0 \ r_c=75^{+85}_{-45} M_{\odot} pc^{-2}$ from \cite{burkert15}({\it grey shadowed area}).}
\end{centering}
\end{figure}

 \begin{figure}
 \begin{centering}
\includegraphics[width=0.8 \textwidth]{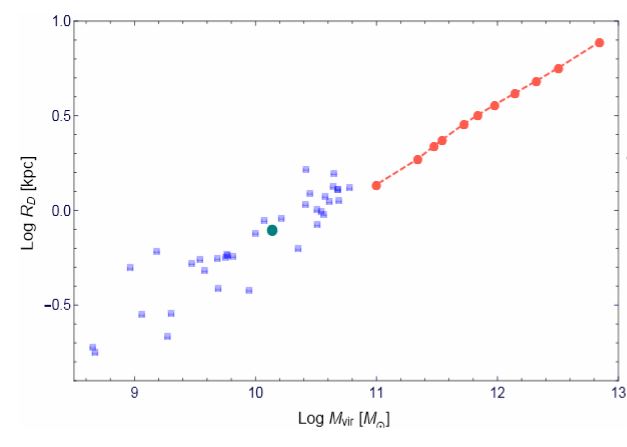}
\vskip 0.1cm
\advance\leftskip 0.7cm
\includegraphics[width=1.0 \textwidth]{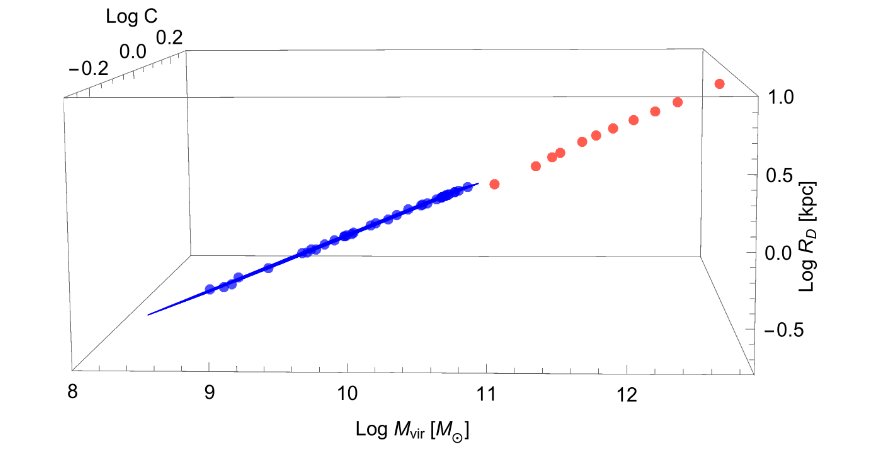}
 \caption{ The the disc lenght scale versus the halo virial mass. 
Red circles represent the (967) spirals, blue squares represent the {\bf dd} ({\it top panel}). The relationship, after
involving the compacteness $log ~ C_\star$ in the {\bf dd} ({\it bottom
panel})}.
\end{centering}
\end{figure}

Also in {\bf dd} the central surface density of the DM haloes proportional to the product $\rho_0 r_0$, is found to be 
constant (see Fig. (13)),
specifically, within 0.25 dex the value found in spirals \cite{donato09}. Noticeably, for both galactic systems
there is not a satisfactory physical explanation for these two observational evidences.

\subsection{\bf A further galaxy structural parameter: the compactness}

In {\bf dd} galaxies, differently from Spirals, the disk length scale $R_{D}$ is not directly related with the virial mass
$M_{vir}$,
see Fig(14). This prevent us from straightforwardly converting, as we do in Spirals, the URC expressed in normalized radius
$x$ to
that expressed in the physical radius $r$ (in fact, in spirals: $r= x\ R_D(M_{vir})$). We can obtain this conversion, by
introducing, for each galaxy of our
Sample, a new observational 
quantity, related to the distribution of stellar disk: the compactness $C_\star$, defined as the ratio between the value of
$R_D$
derived from the disk mass $M_D$ through the regression $ log \ R_D$ vs $ log \ M_D$ found for the whole sample,
and the value $R_D$ directly measured from the photometry. For the Sample of {\bf dd} under study we find:
 $log\ R_D=-3.64+0.46\ log M_D.$ Then,

 \begin{equation}
 log ~C_\star=-3.64+0.46\ log \ M_D -log \ R_D 
\end{equation}

$log ~ C_\star$ defines for the Galaxies of a Sample the differences in the sizes of their stellar discs when all objects are
reduced to a same stellar mass. We
find, with a negligible scatter: 

\begin{equation}
log \ (R_D/kpc)=-4 +0.38\ log ~ (M_{vir}/M_\odot) - 0.94\ log\ C_\star.
\end{equation}
see Fig(14) This result brings two important consequences: i) in {\bf dd} the URC expressed in physical units has two
controlling parameters: the luminosity and the compactness.
ii) it is remarkable and presently unexplained that two secondary properties of the stellar discs of Galaxies, both
belonging to the Luminous World, enter to set a tight relationship with the dark halo mass the most important tag of the
dark world of spirals.

\section{The McGaugh et al 2016 two-acceleration relationship: a challenge for Dark Matter?}

According to recent results (\cite{McG}) fueled by the kinematical and photometric data of 153 spirals, the total
radial
acceleration of the latter,
\begin{equation}
 g\equiv V^2/r 
 \end{equation}
 where $V(r)$ is the circular velocity, shows an anomaly. It correlates, at any radius and in any object, with its
component generated {\it only} from the baryonic matter 
\begin{equation}
g_b \equiv V_b^2/r 
\end{equation}

where $V_b(r)$ is the baryonic contribution 
to $V(r)$ see Fig (15). The baryonic matter, therefore, seems to command the total matter on how to move. This may indicate a
very exotic nature for the Dark particle, or the need for an alternative to the DM paradigm or even, a falsification of the
Galilean Inertia Law \cite{mo3}. In any case, is it that true that: "the (above) relationship appears to be a law of Nature,
a
sort of Kepler's law for rotating galaxies'' (McGaugh et al. (2016)?

In fact, all this must be gauged to the 
phenomenology of the mass distribution of Spirals within the dark matter + Newtonian Gravity paradigm, (see previous
sections).

Let us start with 
\begin{equation}
 g(r)=g_{h}(r)+g_{b}(r)
\end{equation}
 $g(r)$ is the total radial acceleration, while $ g_{h}(r), g_{b}(r)$ are its components generated by the DM halo and by the
baryonic matter, respectively. In detail, at any radius $r$, we have:
 \begin{equation}
 g_b =(V_{d}^{2}+V_{bu}^{2}+V_{HI}^{2})/r =g-g_h, \ \ \ \ \ g_h=V_{h}^{2}/r 
 \end{equation}
 
 where all above quantities are function of galactocentric radius $r$

\begin{figure}
\begin{centering}
\includegraphics[width=1. \textwidth]{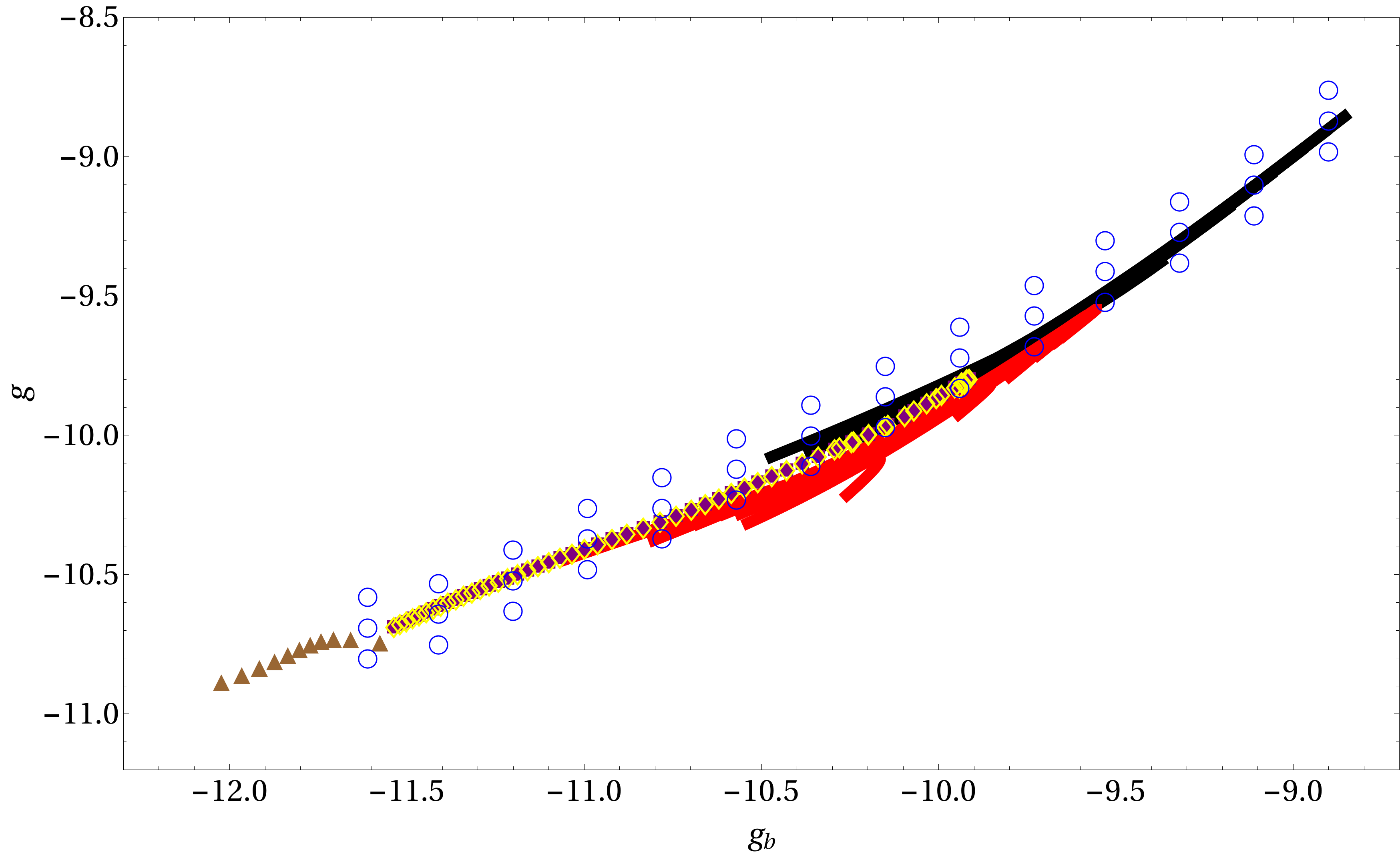}
\caption {The region of the McGaugh et al. 2016 relationship in the $g$-$g_b$ plane including its 1-$\sigma$ uncertainty
({\it blue circles}). Also shown 
the relationships found by \cite{sal2} ({\it red/black
lines, blue squares/yellow diamonds/brown triangles points)}. Accelerations are in units of $Log\ m/s^2$}
\end{centering}
\end{figure}
 
In detail, McGaugh et al. (2016) have investigated a sample of 153 galaxies across a large range of scales in luminosity and
Hubble Types and with high quality rotation curves $V(r)$.
For each object, they derived, at any radius with the RC measurement, the radial acceleration $g(r)$ out to outer galactic
radii and then
compared it with the corresponding value of $g_b(r)$ and the acceleration generated by all the
luminous
matter of the galaxy.
In order to derive the latter accelerations, they used the galaxy surface brightnesses and assumed reasonable values for the
mass/light ratios of the disk and bulge surface/volume densities. Then they inserted them into the relative Poisson
equation. Noticeably, this procedure takes in consideration also stellar disks non perfectly Freeman-like. 

The relationship they found is displayed
as blue circles in Figure (15), extends in $g_b$ for about 3 orders of magnitudes,
with a r.m.s (and a systematics) of 0.11 dex (\cite{McG}). Quantitatively, the relation reads as: 
\begin{equation}
g(r) =g_b(r)/(1-Exp[-(g_b(r)/a_0)^{0.5}])
\end{equation}
 with $a_0=1.2 \times 10^{-10} \ m/s^2$ \cite{McG}. It implies that the baryonic component of the radial acceleration
predicts the latter 
within a 1-$\sigma$ uncertainty of $\pm 13 \% $ (see \cite{McG}). At low $g, g_b$ accelerations, the relation in Eq.(23) and
Fig (15) is clearly very different from that expected in the
Newtonian no Dark Matter framework: $g(r)=g_b(r)$, where the centrifugal acceleration balances the gravitational
acceleration arising from the distribution of all the baryons in the galaxy. 

\cite{sal2} has independently confirmed and statistically extended the results of \cite{McG} by applying 
three different methods, that {\it assume} the presence of DM halos as the origin of no-keplerian features in the
accelerations, to 100000 accelerations measurements from about 1200 spirals (see Fig (13)).

\subsection{\bf The origin of the $g$ vs $g_b$ relationship in Spiral Galaxies}

The latter results conciliate the McGaugh et al relationship and the Dark Matter paradigm. However the decisive step is
to interpret such relationship. For this purpose, we use $V_G(x)$, a model for the circular
velocity of Spirals that hereafter will be called "General". At any $x$, we set 
\begin{equation}
 V_{G}^2(x)= V_{Gh}^2(x) +V_{Gd}^2(x)+ V_{GHI}^2(x)
 \end {equation}

where we adopt the Freeman velocity profile of Eq. (5) for the stellar disk component and Eq. (17) for the HI disk component.
For the dark halo component, we assume: 
\begin{equation}
 V_{Gh}^2(x) = 8.5 \ 10^5 M_D/R_D \ B \ x^{d+1}/(\gamma^2+x^2) \ ( M_D/(10^{11}M_\odot)) ^a \ ( km/s)^2
\end{equation}
with $M_D$ in $M_\odot$ and $R_D$ in kpc. In detail, $B$ is proportional to the fractional
content of dark matter at $R_{opt}$, $ a $ specifies the dependence of the latter quantity on the disk mass, $\gamma$ measures
the size of DM halo core in units of $R_{opt}$ 
 and $d$ indicates how compact is the distribution of dark matter with respect to that of the stars. The disk mass $M_D$ is
the running variable, 
$10^9 ~ M_\odot \leq M_D\ \leq 4\times 10^{11}\ M_{\odot}$, that describes the entire family of spirals. 
 The General halo velocity model in Eq. (24) is very flexible, it
can represent very different DM density profiles, including the NFW and the Burkert ones. We have:

\begin{equation}
g_G(x)= V_{G}^2(x,M_D)/ (x \ R_{opt}) 
\end{equation}
\begin{equation}
\ g_{bG}(x)=(V_{G}^2(x,M_D)-V_{Gh}^2(x,M_D) - V_{GHI}^2(x,M_D))/(x \
R_{opt})
\end{equation}.
 \begin{figure}
\begin{centering}
\includegraphics[width= 1.\textwidth]{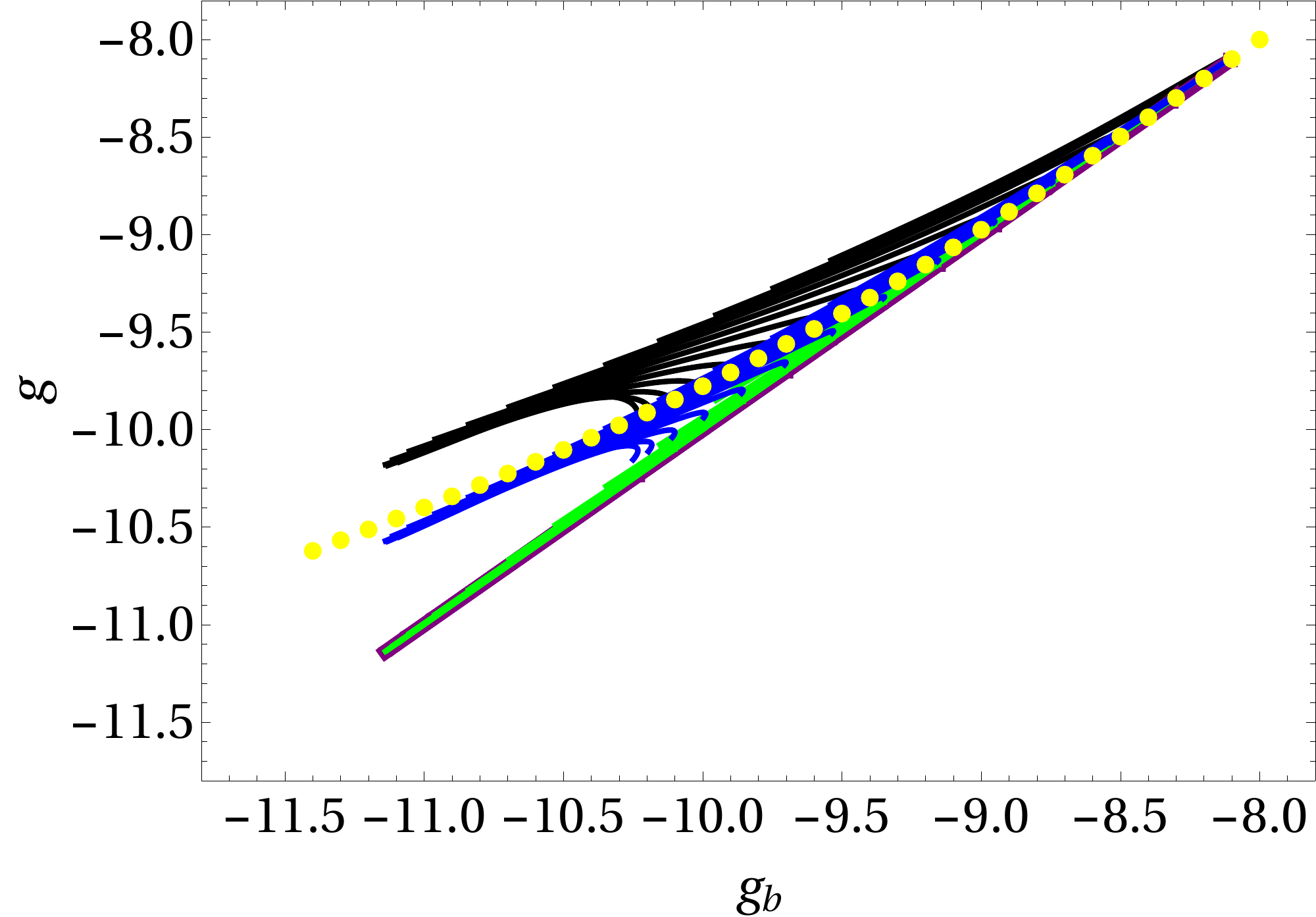}
\caption{The McGaugh et al. (2016) relationship (yellow points) best fitted by the General model (blue thick lines). Also
shown the latter in the cases of -no dark matter (red lines), -compact dark matter (purple lines) - all dominating dark matter
(black lines) and -fraction of DM at $R_{opt}$ increasing with luminosity (green lines).}
\end{centering}
\end{figure}

We use the General model and the relationship in Eq. (4) to fit the data of the McGaugh et al. (2016) relationship (see Fig
(16)).
The fit is excellent and the best fit values are 

$$
\gamma =1, \,\, a=-1/2, \,\, d=2, \,\, B=0.1
$$

 (see Fig(16)). The fitting uncertainties of these
parameters are about 15\%. Let us now determine the cases in which the accelerations from the General model 
{\it fail} to recover the McGaugh et al. (2016)
relationship. The parameter $\gamma$ plays little role in the agreement between these two relationships: we can take $0.4
<\alpha<\infty $ without breaking it. The McGaugh, 2016 relationship is blind to the inner distribution of dark matter.
Instead, for values $B\simeq 0 $ (no dark matter) or $B>0.3$ which corresponds to an amount of DM $ > 3$
times the best fit value, the General model fails to reproduce 
McGaugh et al. (2016) relationship (see Fig (15)). Similarly, the agreement between the relationships continues
 for values of $d$ different from the best fit value of 2, but, for $d<-2/3$, i.e. for a DM halo more
compact than the
luminous matter, the agreement breaks down. Finally, the agreement breaks down if the quantity $c> 0$ (see Fig(16)) that
indicates that, at any $x$, there must be a larger fraction of DM in the higher luminosity objects. 

Therefore, the McGaugh et al. (2016) relationship exists
because and only because, in spirals, dark and luminous matter are distributed in the following way:

$\bullet$ {\it i)} in every object the luminous matter is more concentrated than the dark matter: the quantity $g_h(r)/g_b(r)$
increases with radius $r$ 

$\bullet$ {\it ii)} at any fixed radius $x$, the lower is the luminosity of the object, the larger is the fraction of dark
matter: the quantity $g_h(x,M_I)/g_b(x,M_I)$ increases with decreasing galaxy luminosity.

It is easy to show that {\it i)} and {\it ii)} lead to the above $g(g_b)$ relationship They are known long since to arise from
well known astrophysics (\cite{ps88}, 
\cite{ash}). Evidence {\it i)} originates from the fact
that the dark particles are virtually collisionless with respect to the baryonic particles ( e.g.
\cite{sod,wei}, \cite{dek}). Evidence {\it ii)} is related to the fact that the smaller the gravitational potential
well is, the more efficiently the energy injected into the interstellar space by Supernovae
explosions can remove the neutral hydrogen of the galaxies, preventing it to be turned into stars. (e.g. \cite{dek})
It is worth to mention that {\it i)}, and {\it ii)} are found in the Hydro N-Body simulations performed in
$\Lambda CDM$ scenario e.g. \cite{sod} 
 
 The paradigm of halo dark matter around spirals is therefore totally compatible with the McGaugh et al. (2016) relationship,
that, in turn, acquires a physical explanation. The McGaugh relationship does not 
challenges the $\Lambda$CDM scenario. In fact, as a confirm, a very similar relationships directly emerges in a set 
of well-resolved galaxies in the EAGLE suite of $\Lambda CDM$ hydrodynamic simulations \cite{let}.

\section{Conclusions}

 Thirty years of investigations have secured the evidence of a dark component around the disk of galaxies of any luminosity.
The rotation curves of disk systems, excellent tracers of their gravitational fields, are described by an universal profile
$V_{URC}(r/R_{opt}, M_I)$ which is incompatible with their distribution of star and gas. Noticeably, the URC,
although amply dominated by the Dark Matter component, is a function of {\it a)} the radius in units of disk lenght-scale
$R_D=1/3.2 \ 
R_{opt}$, {\it b)} the magnitude $M_I$, and {\it c)} the stellar concentration $C$, all quantities of the luminous component.
This is extremely
remarkable and it could indicate a non-standard nature of the dark matter. 

Furthermore, by modelling the URC with standard disk + halo components, we find that the three parameters of the velocity
model: the disk mass $M_D$, the DM core radius $r_0$ and the central density $\rho_0$ are all interrelated to each other and
to the galaxy luminosity. This result, in connection with the cored density profiles routinely found in spirals, seems to
be at variance with the paradigm of collisionless dark matter and to indicate us that the distribution of matter in galaxies
might
be a portal for new physics. 

The main alternative (not discussed in the work) to this change of paradigm is to upgrade the baryonic components to a
crucial role during the period of the formation of spiral's disks (e.g. \cite{dicintio14,t13}). In this scenario it is 
proposed that stars, when go supernovae,
can transfer their original nuclear energy to the kinetical energy of the DM particles. This process, could modify a cusped DM
halo density distribution into one with a flat inner core and, in addition, it could create the ensemble of the relationships
among the halo and disk structural parameters, found in Spirals.

 The URC plays also a decisive role in the investigation on the recent claim, raised by McGaugh et al. 2016, of a further
challenge to the paradigm of Newtonian DM halos based on the finding of a tight correlation, at any galactic radius of any
spiral, between the radial acceleration $g(r)$ and its baryonic component $g_b(r)$. The URC, in fact, confirms the existence
of such
relation in normal Spirals, but, at the same time, shows that this relation exists also within the standard DM halo paradigm
and it has simple physical explanations.

.

%\begin{figure}
% \includegraphics{example.eps}
%label{fig:1} % Give a unique label
%\end{figure}
%\begin{figure*}
 % \includegraphics[width=0.75\textwidth]{example.eps}
%\label{fig:2} % Give a unique label
%\end{figure*}
%

\begin{acknowledgements}

I would like to thank the Specola Vaticana and the Organizers of the Workshop ``Black Holes, Gravitational Waves and
Spacetime Singularities'' where this paper has been conceived and Gabriele Gionti for useful discussions.
\end{acknowledgements}

%\bibliographystyle{spbasic} % basic style, author-year citations
%\bibliographystyle{spmpsci} % mathematics and physical sciences
%\bibliographystyle{spphys} % APS-like style for physics
%\bibliography{} % name your BibTeX data base

 

\end{document}